\documentclass[pdftex,twocolumn,epjc3]{svjour3}          

\RequirePackage[T1]{fontenc}

\smartqed  

\RequirePackage{graphicx}
\RequirePackage{mathptmx}      
\RequirePackage{flushend}
\RequirePackage[numbers,sort&compress]{natbib}
\RequirePackage[colorlinks,citecolor=blue,urlcolor=blue,linkcolor=blue]{hyperref}

\usepackage{amsmath,amsfonts}
\usepackage{bbm}

\newcommand{\tr}{\mathop{\mathrm{tr}}}

\journalname{Eur. Phys. J. C}

\begin{document}

\title{Interpretation of bulk viscosity as the generalized Chaplygin gas}



\author{
Marek Szyd{\l}owski\thanksref{addr1,addr2,e1}
\and
Adam Krawiec\thanksref{addr3,e2}
}

\thankstext{e1}{e-mail: marek.szydlowski@uj.edu.pl}
\thankstext{e2}{e-mail: adam.krawiec@uj.edu.pl}

\authorrunning{} 

\institute{Astronomical Observatory, Jagiellonian University, Orla 171, 30-244 Krak{\'o}w, Poland \label{addr1}
\and
Mark Kac Complex Systems Research Centre, Jagiellonian University, {\L}ojasiewicza 11, 30-348 Krak{\'o}w, Poland \label{addr2}
\and
Institute of Economics, Finance and Management, Jagiellonian University, {\L}ojasiewicza 4, 30-348 Krak{\'o}w, Poland \label{addr3}
}

\date{Received: date / Accepted: date}

\maketitle

\begin{abstract}
The cosmological observations suggest that the presently accelerating universe should be filled by an exotic form of matter, violating the strong energy condition, of unknown nature and origin. We propose the viscous dark matter of a source of acceleration in the form of Chaplygin gas which is characterized by equation of state in the phenomenological form $p=-\frac{A}{\rho^{\alpha}}$, where $p$ and $\rho$ are pressure and energy density respectively ($A$ and $\alpha$ are constants). Chaplygin gas is interpreted in terms of viscous matter and without the cosmological constant. The acceleration effect is caused only by viscosity in this class of cosmological models. We show that bulk viscosity effects introduced to the standard FRW cosmology give rise to the natural unification of both dark matter and dark energy. We show that dust viscous cosmological models are structurally stable if $m < 1/2$ ($1+\alpha=1/2-m$).
\end{abstract}

\section{Introduction}

In the pursuit of solving the Universe acceleration problem, different candidates for the description of dark energy were proposed. One of them was fluid called Chaplygin gas and described by the equation of state in the phenomenological form $p=-\frac{A}{\rho^{\alpha}}$, where $p$ and $\rho$ are pressure and energy density respectively ($A$ and $\alpha$ are constants). The problem of interpretation of this equation of state was addressed by Gorini et al. \cite{Gorini:2004by} who pointed out the importance of a possible theoretical basis for the Chaplygin gas in cosmology. They were looking for a fundamental mechanism to produce this form of fluid as a source term in the right-hand side of the Einstein equation. As a proposition of such a mechanism they indicated the brane model. Our investigation are going to find another possible theoretical basis for Chaplygin gas. Therefore, we shift our attention from dark energy to dark matter to investigate the cosmology with fluid described by the equation of state for Chaplygin gas.

In their pioneering papers, both Murphy and Klimek investigated the dynamics of cosmological models with bulk viscosity \cite{Murphy:1973zz,Klimek:1981vf}. Klimek studied models with Einstein's (positive and negative) cosmological constant and viscosity coefficient being the power function of matter density with the exponent $m$ ($\xi(\rho) = \beta \rho^m$). He found some interesting phase space structures for some positive values of parameter $m$. While he found some exotic dynamics in some specific intervals of positive values of exponent parameter, he left the domain of negative values of the parameter $m$ unexplored. The present state of research on the viscous cosmology see \cite{Brevik:2017msy}.

In general, in the past studies of bulk viscosity cases of $m < 0$ were neglected. In this paper we consider the viscosity as a property of Chaplygin gas and study the dynamics of the cosmological model with viscous fluid (Chaplygin gas) and baryonic matter (dust). It can be shown that matter with bulk viscosity described by the parameter $m$ in the interval $[-\frac{3}{2}, -\frac{1}{2})$ corresponds to matter in form of (generalized) Chaplygin gas with parameter $\alpha \in (0,1]$. Moreover, this cosmological model with (generalized) Chaplygin gas is supported by present astronomical data. Therefore, cosmological model with bulk viscosity requires more scrupulous studies.

Taking the equation of state in the form $p = B\rho - A/\rho^{\alpha}$ we obtain the model with modified Chaplygin gas \cite{Benaoum:2002zs,Debnath:2004cd,Bento:2004uh}. Brevik and Gorbunova considered this model with the parameter $\alpha < -1/2$ and found that the future singularity $t=t_s$ is affected from bulk viscosity \cite{Brevik:2008xv}. Benaoum presented the review for this model \cite{Benaoum:2012uk}. Fabris et al. found that this extension is ruled out as power spectrum observational data support $B=0$ and reduction to the generalized Chaplygin gas model \cite{Fabris:2010vd}.

The equation of state in the form $p=-\frac{A}{\rho}$ was first put forward in 1904 by the Russian physicist S.~A. Chaplygin to describe the adiabatic process in the aerodynamical context \cite{Chaplygin:1904gj,Kamenshchik:2001cp}. This form of equation of state was recently proposed as a candidate for dark energy in the Universe \cite{Gorini:2002kf}. Although the postulated form of equation of state has purely phenomenological character, currently its generalizations admitting the more general form of equation of state $p=-\frac{A}{{\rho}^{\alpha}}$ where $0\leq \alpha \leq 1$ have been proposed \cite{Bento:2002ps}. It was pointed out by Bento et al. that the model with Chaplygin gas can be described by a complex scalar field which action can be written as the generalized Born--Infeld action. Attractiveness of the equation of state for Chaplygin gas comes from the fact that it gives the unification of dark energy and also dark matter which is proposed in astrophysics to explain the flat rotation curves of spiral galaxies \cite{Bilic:2001cg}. In cosmological models with Chaplygin gas the problem of inflationary dynamics was discussed \cite{Herrera:2016sov,Barrow:2016qkh}. Many authors have studied constraints on the model parameters from various observational data. The stringent constraint on the Chaplygin gas can be obtained from CMB anisotropy measurements \cite{Bento:2002yx}, the gravitational lensing surveys \cite{Dev:2002qa}, the X-ray gas fractions of clusters \cite{Cunha:2003vg,Zhu:2004aq} the age measurements of high-$z$ objects \cite{Alcaniz:2002yt}, SNIa \cite{Fabris:2002vu,Biesiada:2004td,Xu:2012qx}, the global 21-cm absorption signal \cite{Yang:2019nhz,Pigozzo:2019koi} as well as the Bayesian analysis of multi-data \cite{Sasidharan:2018bay,Herrera-Zamorano:2020rdh}. This constraint in terms of two parameters of Chaplygin gas $(A_{s},\alpha)$ is
\begin{equation}\label{eq:1}
\rho_{\text{Chapl}}=\rho_{\text{Chapl,0}} \left[ A_{s}+(1-A_{s})a^{-3(1+\alpha)} \right]^{\frac{1}{1+\alpha}}
\end{equation}
where $\rho_{\text{Chapl,0}}$ is energy density of Chaplygin gas at present, $A_{s}=A/\rho^{1+\alpha}_{\text{Chapl,0}}$ is substitution of the parameter $A$. The relation (\ref{eq:1}) is a consequence of conservation condition
\begin{equation}\label{eq:2}
\dot{\rho}+3\frac{\dot{a}}{a}(\rho+p)=0
\end{equation}
The both parameters of the model with Chaplygin gas are related
\begin{align}
&\frac{dp}{d\rho} =-\frac{\alpha p}{\rho} \label{eq:3} \\
&\frac{dp}{d\rho_{\text{Chapl,0}}} = \alpha A_{s}\rho_{\text{Chapl,0}}^{\alpha-1}. \label{eq:4}
\end{align}
This derivative is positive provided that $\alpha > 0$. This quantity should not be interpreted as a square of speed of sound in this fluid. Bulk viscous pressure $\Pi$ is defined as $\Pi= - \xi \Theta$, where $\Theta$ is the expansion scalar $\Theta = u^{\mu}_{;\mu}$. So, it depends on the four-velocity divergence. When one perturbs this quantity in a covariant way the effective speed of sound of the bulk viscous fluid has a different structure, i.e. other than the mere adiabatic convention $\partial p /\partial \rho$.

Let us consider now the standard FRW model filled with some dissipative fluid. If we introduce bulk viscosity to the FRW models then conservation condition has the form
\begin{equation}\label{eq:6}
\dot{\rho}=-3H [ \rho+(\gamma \rho-3\xi(\rho)H) ]
\end{equation}
Several cosmological models have been studied in which the fluid has the bulk viscosity coefficient, usually parameterized, following Belinskii and Khalatnikov, in a power law
\begin{equation}\label{eq:5}
\xi(\rho) = \beta \rho^{m}
\end{equation}
where $\beta$ and $m$ are constants \cite{Belinskii:1975iv,Belinskii:1979ic,Chimento:1993zc,vanElst:1994bn,Pavon:1990qf,Klimek:1981vf}.

Conservation condition (\ref{eq:6}) with the Belinskii--Khalatnikov parameterization viscosity coefficient (\ref{eq:5}) can be integrated and the solutions are
\begin{equation}\label{eq:7}
\rho(a)= \begin{cases}
\rho_{0}a^{-3(1+\gamma-\bar{\alpha})},\quad & m =1/2 \quad (\alpha=-1) \\
\Big( A+\frac{B}{a^{3(\frac{1}{2}-m)(1+\gamma)}} \Big)^{\frac{1}{\frac{1}{2}-m}} \quad & m \ne 1/2 \quad (\alpha \ne -1)
\end{cases}
\end{equation}
where $\bar{\alpha}=\beta \sqrt{3}$.

In further consideration we ignore an exceptional case of $m=1/2$. This case corresponds to a situation the equation of state of Chaplygin gas assumes the form of perfect fluid with the equation of state $p = - \bar{\alpha}\rho$.

\section{Bulk viscosity interpretation as generalized Chaplygin gas}

Let $k=0$ (flat model) filled with dissipative fluid with pressure and non-interacting matter with the equation of state $p = \gamma \rho$. Then $\rho(a)$ can be integrated in an exact form from (\ref{eq:6})
\begin{align}
p &= \gamma \rho-3\xi(\rho)H \label{eq:8}\\
\rho &= 3H^{2} \label{eq:9}
\end{align}
\begin{equation}\label{eq:10}
\frac{d\rho}{d a}=-\frac{3\rho}{a} \left[ (1+\gamma) - \sqrt{3} \beta \rho^{m-\frac{1}{2}} \right]
\end{equation}

We consider an expanding universe and choose $H(t)$ to be positive. It would be convenient to introduce a new variable $E$ instead of the original variable $\rho$, namely
\[
E \colon \rho^{\frac{1}{2}-m} = E.
\]
We rewritten the conservation condition to the new form in terms of elasticity of $\rho$ with respect to the scale factor $a$
\[
\frac{1}{3} \frac{d\ln \rho}{d\ln E} \frac{d\ln E}{d\ln a} = (1 + \gamma) - \sqrt{3} \beta \rho^{m - \frac{1}{2}}.
\]
Then after introducing a new Hubble time parameter $\tau$
\[
\tau=\ln a(t)
\]
we obtain a linear equation in the form
\[
\frac{dE}{d\tau} = -3\left(\frac{1}{2} - m\right)(1 + \gamma)E + 3 \sqrt{3} \beta.
\]
Finally as a solution we obtain
\begin{align*}
E(\tau) &= B \exp\left[-3\left(\frac{1}{2}-m\right)(1 + \gamma)\right]\tau + \frac{\sqrt{3} \beta}{(\frac{1}{2}-m)(1 + \gamma)} \\ &= B \exp\left[-3\left(\frac{1}{2}-m\right)(1 + \gamma)\right]\tau + A,
\end{align*}
where $A = \sqrt{3} \beta /\left(\frac{1}{2}-m\right)(1 + \gamma)$ is positive if $\left(\frac{1}{2}-m\right)(1 + \gamma)$ is positive.

Therefore $E(a)$ dependence has the form
\[
E(a) = Ba^{-3\left(\frac{1}{2}-m\right)(1 + \gamma)} + A
\]
or
\begin{equation} \label{eq:11}
\rho(a) = \left[Ba^{-3\left(\frac{1}{2}-m\right)(1 + \gamma)} + A \right]^{\frac{1}{\frac{1}{2}-m}}.
\end{equation}
In consequence
\[
V(a) = - \frac{1}{6} a^2 \left[Ba^{-3\left(\frac{1}{2}-m\right)(1 + \gamma)} + A \right]^{\frac{1}{\frac{1}{2}-m}}.
\]

In our further analysis we assume that $A$ is positive and will parameterize constant $A$ and $B$ in terms of the parameter $A_s$.

Let us consider Chaplygin gas for comparison
\begin{equation}\label{eq:12}
\rho(a)=\Big( A+\frac{B}{a^{3(1+\alpha)}} \Big)^{\frac{1}{1+\alpha}}
\end{equation}
where comparison (\ref{eq:11}) and (\ref{eq:12}) gives $1+\alpha=(1+\gamma)(\frac{1}{2}-m)$. Then for $\gamma=0$ (dust) we obtain
\begin{equation}\label{eq:13}
\alpha=-\frac{1}{2}-m.
\end{equation}
Therefore, for dust matter with viscosity parameterized by the Belinskii--Khalatnikov viscosity coefficient (\ref{eq:5}) we obtain
\begin{equation}\label{eq:14}
\rho(a)=\Big[ A+\frac{B}{a^{3(1+\gamma)(1+\alpha)}} \Big]^{\frac{1}{1+\alpha}}.
\end{equation}
By choosing both positive values of $A$ and $B$ constants we see that for small $a$ and positive $\alpha$ the above expression can be approximated by
\begin{equation}\label{eq:15}
\rho \propto \frac{B^{\frac{1}{1+\alpha}}}{a^{3(1+\gamma)}}
\end{equation}
that corresponds the universe dominated by perfect fluid scaling like $\rho \propto a^{-3(1+\gamma)}$. For large value of scale factor $a$, if $\alpha$ is positive then
\begin{equation}\label{eq:16}
\rho \propto A^{\frac{1}{1+\alpha}}
\end{equation}
which, in turn corresponds to an empty universe with the cosmological constant
\begin{equation}\label{eq:17}
\Lambda=A^{\frac{1}{1+\alpha}}
\end{equation}
Note that in opposite case if $\alpha<0$ is considered then $\rho(a)$ relation interpolates between cosmological constant and matter dominated phases of evolution. If $\gamma=0$ (dust) is assumed then our model interpolates between the dust dominated phase and de Sitter phase. As an intermediate regime we obtain a mixture of dust and cosmological term. In general case it is a regime described by the equation of state for some mixture $p=\alpha \rho$ and the cosmological constant fluid.

The universe is accelerating provided that
\begin{equation}\label{eq:18}
\rho+3p=\rho-\frac{3A}{\rho^{\alpha}}<0\, \rightarrow \,
\rho<(3A)^{\frac{1}{1+\alpha}}.
\end{equation}
The condition $\ddot{a}>0$ is equivalent to
\begin{equation}\label{eq:19}
a^{3(1+\gamma)(\frac{1}{2}-m)}>\frac{B}{2A}
\end{equation}

Due to the Hamiltonian approach to the quintessential cosmology with general form of equation of state $p=w(a)\rho$ it can be useful to represent it in the form of a Hamiltonian dynamical system
\begin{equation}\label{eq:20}
\mathcal{H}=\frac{\dot{a}^{2}}{2}+V(a)=-\frac{k}{2}
\end{equation}
where $V(a)=-\frac{\rho_{\text{eff}}a^2}{6} = - \frac{a^2}{6} \sum_i \rho_i = \sum_i V_i$ and $k$ is the scalar curvature and motion of the system takes place over energy levels $E=-k/2=\text{const}$. In our further analysis we consider the case $k=0$.

The corresponding Hamiltonian system is
\begin{align}
& \dot{x} = y, \label{eq:21}\\
& \dot{y} = -\frac{\partial V}{\partial x}, \label{eq:22}\\
& \frac{y^{2}}{2}+V(x) = \frac{\Omega_{\text{k,0}}}{2},\label{eq:23}
\end{align}
where $x$ is the value of the scale factor expressed in the units of present value and
\begin{equation}\label{eq:24}
V(x)=-\frac{1}{2}\Omega_{x}(x)x^{2}-\frac{1}{2}\Omega_{\text{b,0}}x^{-1}
\end{equation}
here the new variable $\tau$ is the reparameterized time variable 
\[
t\rightarrow \tau \colon H_{0}dt=d\tau.
\]
The potential (\ref{eq:24}) contains the contribution from Chaplygin gas with the density parameter $\Omega_{x,0}$
\begin{equation}\label{eq:25}
\Omega_{x}=\Omega_{x,0} \left[ A_{s}+\frac{1-A_{s}}{x^{3(1+\gamma)(\frac{1}{2}-m)}} \right]^{\frac{1}{\frac{1}{2}-m}}
\end{equation}
where $\Omega_{x,0}$ is the density parameter for viscous matter with the equation of state $p = \gamma \rho$ and the contribution from baryonic pressureless matter with the density parameter $\Omega_\text{b,0}$.

Subsequently, from equation (\ref{eq:23}) for $\dot{x}=1$ ($H=H_{0}$) we obtain the constraint
\begin{equation}\label{eq:26}
\Omega_{x,0}+\Omega_\text{b,0}=1.
\end{equation}

\begin{figure}
\includegraphics[width=0.5\textwidth]{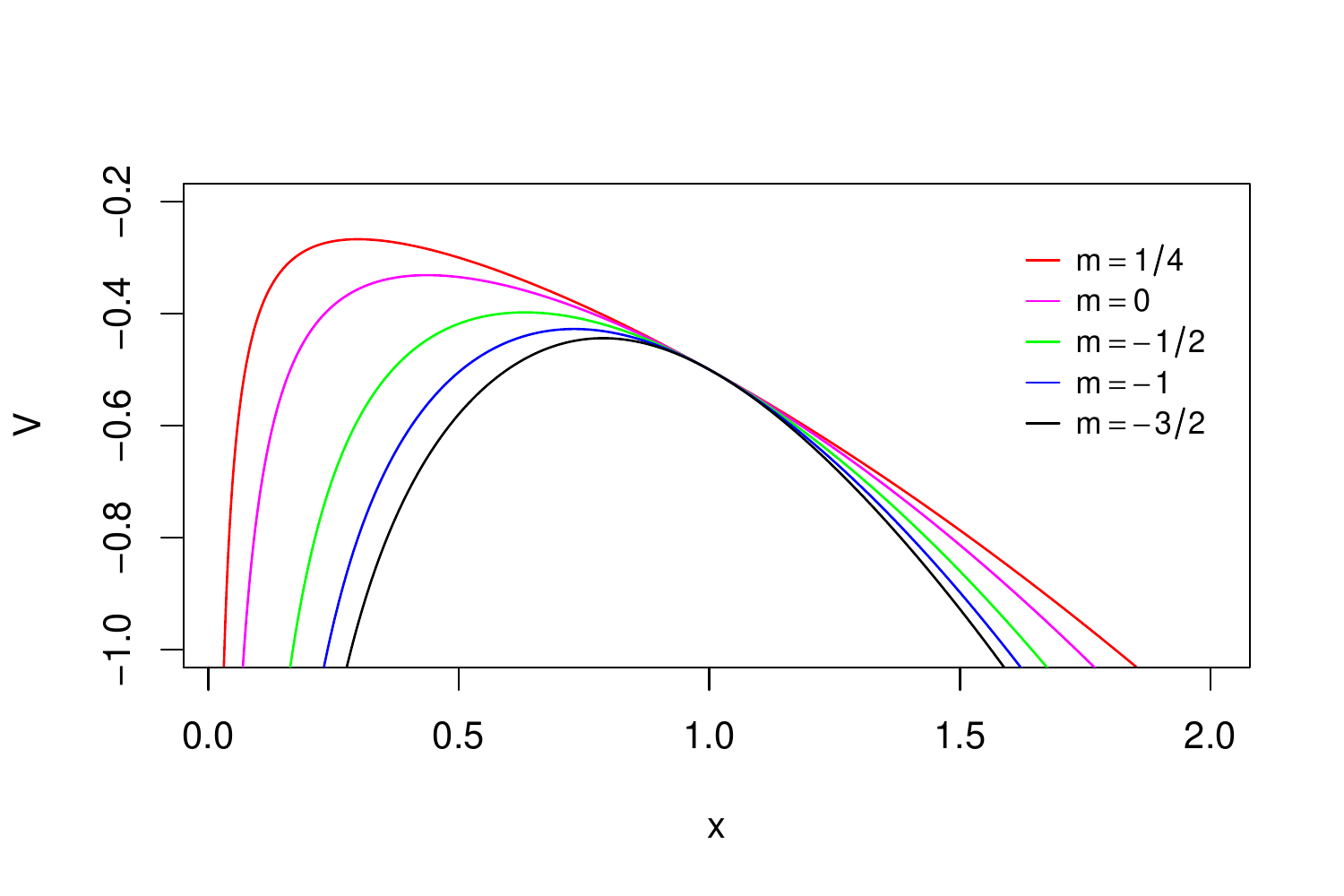}
\caption{The potential (\ref{eq:24}) with $A_s=0.7$, $\Omega_\text{b,0}=0.05$, $\Omega_{x,0}=0.95$ for five values of $m$ in the interval $m < \frac{1}{2}$: the cosmological constant ($m=-1/2$), generalized Chaplygin gas (chosen $m=-1$ from the interval $(-1/2,-3/2)$) and Chaplygin gas ($m=-3/2$). The deceleration phase is for $a < a_\text{saddle}$ and the acceleration phase is for $a > a_\text{saddle}$. The domain below the diagram of potential function is forbidden for the classical motion from the condition $y^2 \ge 0$. \label{fig:1}}
\end{figure}

\begin{figure}
\includegraphics[width=0.5\textwidth]{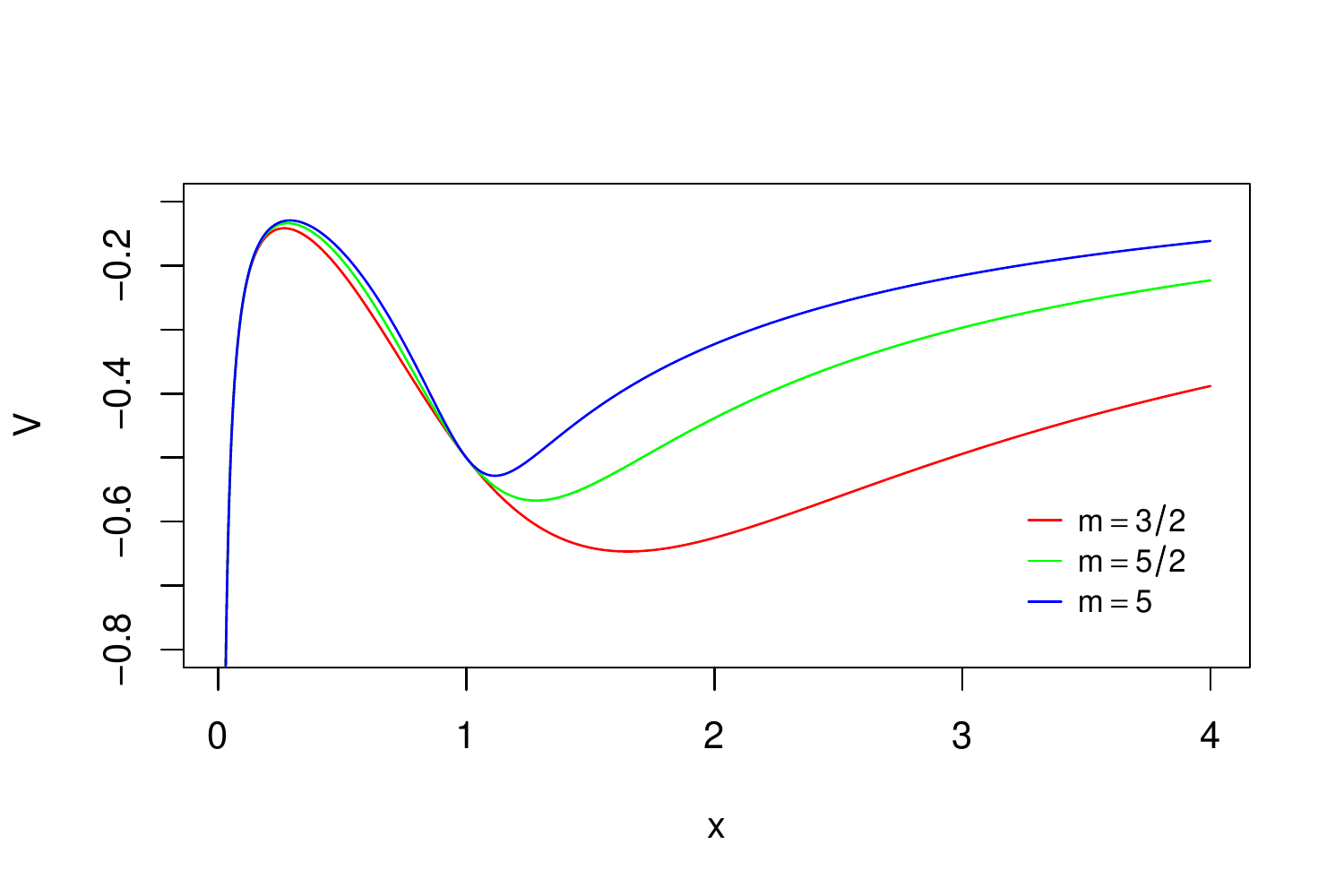}
\caption{The potential (\ref{eq:24}) with $A_s=0.7$, $\Omega_\text{b,0}=0.05$, $\Omega_{x,0}=0.95$ for three values of $m$ in the interval $m > \frac{1}{2}$. The deceleration phase is $a < a_\text{saddle}$ and $a > a_\text{center}$ while the acceleration phase is for $a_\text{saddle} < a < a_\text{center}$. The domain below the diagram of potential function is forbidden for the classical motion from the condition $y^2 \ge 0$. \label{fig:2}}
\end{figure}

The potential (\ref{eq:24}) depends on the parameter $m$. We consider two qualitatively different cases: $m < \frac{1}{2}$ and $m > \frac{1}{2}$. The former is presented in figure~\ref{fig:1} and the latter in figure~\ref{fig:2}. A maximum and minimum on the diagrams of potential functions correspond to a saddle and center, respectively, on a phase portrait. Given the acceleration equation, the possibility of the universe acceleration can be determined from a shape of the potential function. The deceleration phase corresponds to the increasing function $V(a)$ and the acceleration phase to the decreasing function $V(a)$.

\begin{align}
\dot{x} &= y \label{eq:27}\\
\dot{y} &= \frac{\partial}{\partial x} \left[\frac{1}{2}\Omega_\text{b,0}x^{-1}+\frac{1}{2}\Omega_{x,0}x^{2}
\left( A_{s}+\frac{1-A_{s}}{x^{3(1+\gamma)(\frac{1}{2}-m))}} \right)^{\frac{1}{\frac{1}{2}-m}} \right] \label{eq:28}
\end{align}
with the first integral
\begin{equation}\label{eq:29}
\frac{y^{2}}{2} = \frac{1}{2}x^{2}\Omega_{x,0} \Big( A_{s}+\frac{1-A_{s}}
{x^{3(1+\gamma)(\frac{1}{2}-m)}} \Big)^{\frac{1}{\frac{1}{2}-m}}+\frac{1}{2}\Omega_\text{b,0}x^{-1}.
\end{equation}

Let us note that phase space is invariant with respect to mirror $y \to -y$ and $x \to x$.

This relation represent algebraic equation for phase trajectories in phase space $(x,y)$. The evolution of the scale factor is represented by zero energy level in fig.~\ref{fig:1} because we assume $k=0$. The domain beneath the parabola is forbidden for the classical motion if we assume $k=0$.

In the recent series of papers the quantum effects of massive scalar fields are considered on the background of FRW model. These corrections reproduce equation of state for the viscous radiation fluid which correspond $\gamma=\frac{1}{3}$ and then
\begin{equation}\label{eq:30}
\rho (a)=\left( A+\frac{B}{a^{4\left(\frac{1}{2}-m\right)}} \right)^{\frac{1}{\frac{1}{2}-m}}
\end{equation}
Therefore, this form of the equation of state interpolates two different phases of evolution of the universe, namely, the phase of radiation domination and the phase of de Sitter through an intermediate phase of coexistence both effects.

It is possible to find the exact solution for the model with viscous fluid in the form of the following dependence
\begin{multline} \label{eq:31}
t(a) = \frac{2}{\sqrt{3}} (1-A_s)^{\frac{-1}{1-2m}} a^{\frac{3}{2}} \times \\
\times {}_2 F_1 \left( \frac{1}{1-2m}, \frac{1}{1-2m}, \frac{2-2m}{1-2m}, -\frac{A_s}{1-A_s}a^{3\left( \frac{1}{2}-m \right)} \right).
\end{multline}

Kamenshchik et al. found the exact solution for the flat case of cosmological model with the Chaplygin gas \cite{Kamenshchik:2001cp}
\[
t(a) = \frac{1}{6\sqrt[4]{A}} \left( \ln \frac{\sqrt[4]{A+\frac{B}{a^6}} + \sqrt[4]{A}}{\sqrt[4]{A+\frac{B}{a^6}} - \sqrt[4]{A}} - 2 \arctan \sqrt[4]{1 + \frac{B}{A a^6}} \right).
\]
which is a special case of (\ref{eq:31}) for $A=A_s$, $B=1-A_s$ and $m=-3/2$ ($\alpha = 1$).

\section{Dynamical analysis}

The cosmological models with Chaplygin gas were tested empirically with astronomical data in many papers \cite{Cunha:2003vg,Zhu:2004aq,Biesiada:2004td,Lu:2008zzb,Salahedin:2020emw}. The stability of the FRW models with Chaplygin gas was investigated by Szyd{\l}owski and Czaja \cite{Szydlowski:2003wx}.

In this context, using the Akaike information criterion (AIC) of model selection for model comparison, it is found that recent observational data support the MCG model as well as other popular models. The statistical analysis of astronomical data gave the best fit value of the three parameters $(\gamma, A_s, \alpha)$ in the MCG model as $(-0.085, 0.822, 1.724)$ \cite{Szydlowski:2006ay}.

Estimation gives for example that $A_s = 0.70_{-0.17}^{+0.16}$ and $\alpha = -0.09_{-0.33}^{+0.54}$, at a $95\%$ confidence level, which is consistent within the errors with the standard dark matter + dark energy model, i.e., the case of $\alpha=0$. Particularly, the standard Chaplygin gas ($\alpha=1$) is ruled out as a feasible UDME by the data at a $99\%$ confidence level \cite{Zhu:2004aq}.

In the standard cosmological model ($\Lambda$CDM model), both dark and visible matter are pressureless. To investigate the role of baryonic matter in the viscous cosmological model, the two cases should be considered
\begin{enumerate}
\item all matter (dark and visible) is viscous;
\item dark matter is viscous but baryonic matter is non-viscous, treated as an additional non-interacting fluid.
\end{enumerate}

From the dynamical analysis we find that dynamical evolution in both cases is equivalent quantitatively. Therefore, the following dynamical analysis targets the second case.

\begin{figure}
\includegraphics[width=0.5\textwidth]{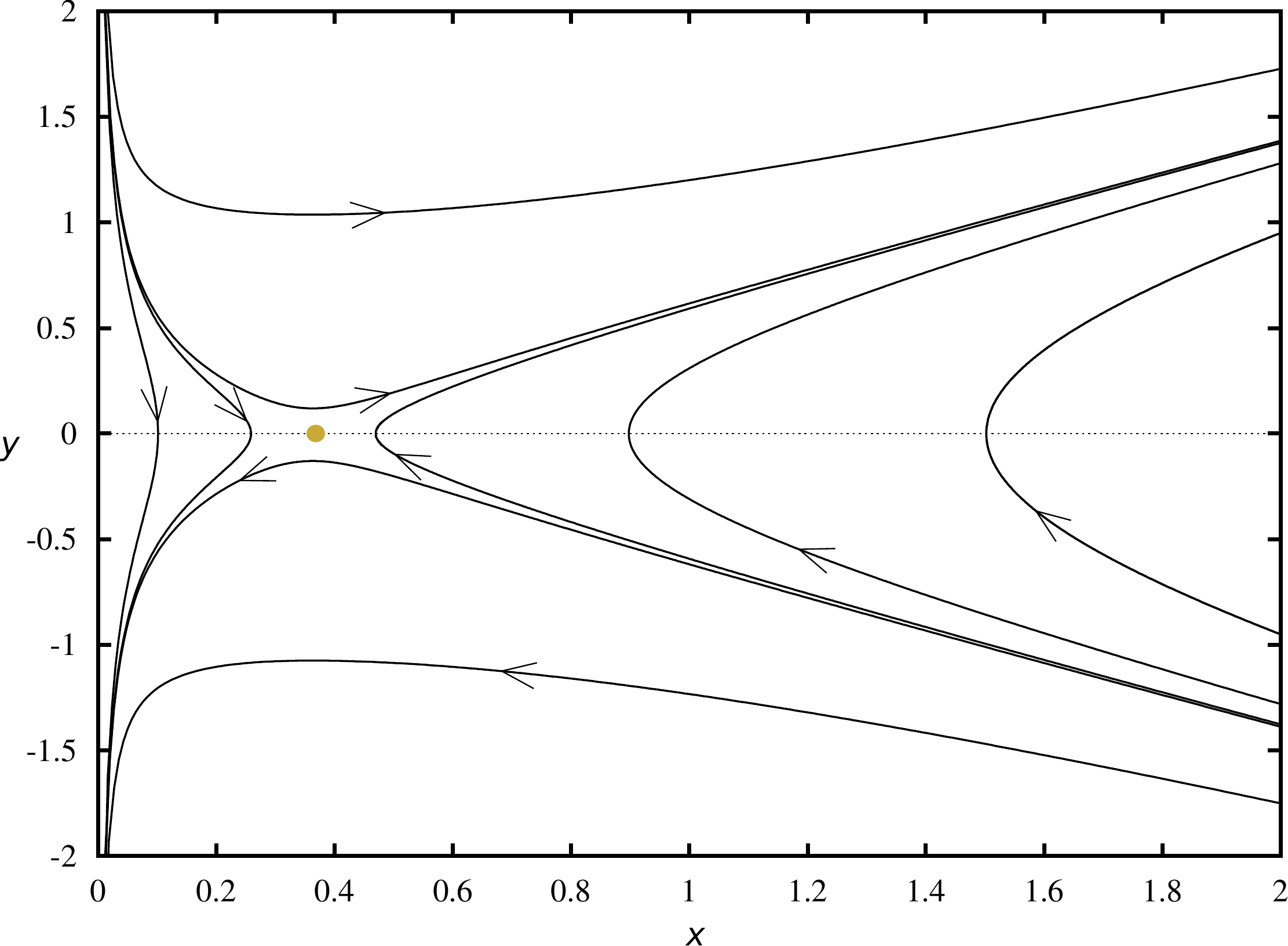}
\caption{The phase portrait of the cosmological model (\ref{eq:27})-(\ref{eq:28}) with $A_s=0.7$, $\Omega_\text{b,0}=0.05$, $\Omega_{x,0}=0.95$ and Chaplygin gas $m=-3/2$ ($\alpha = 1$). Trajectories on the right to the saddle are for the accelerating universe. \label{fig:3}}
\end{figure}

In the dynamical analysis we assume that $A_s = 0.7$, $\gamma =0$, $\Omega_{x,0}=0.95$ and $\Omega_\text{b,0}=0.05$ (baryonic matter). Then we study qualitatively the dynamics of cosmological model (\ref{eq:27})-(\ref{eq:28}) with the parameter $m$ chosen both from interval $m < \frac{1}{2}$ and $m < \frac{1}{2}$ (or equivalently $\alpha > -1 $ and $\alpha < -1$). We draw their phase portraits using XPPAUT software \cite{Ermentrout:2002sa}. In figure~\ref{fig:3} the phase portrait of the cosmological model with Chaplygin gas $m=-3/2$ is presented. Note that for large $x$ trajectories focus near a line $y/x = \text{const}$, they go asymptotically to the de Sitter universe.

\begin{figure}
\includegraphics[width=0.5\textwidth]{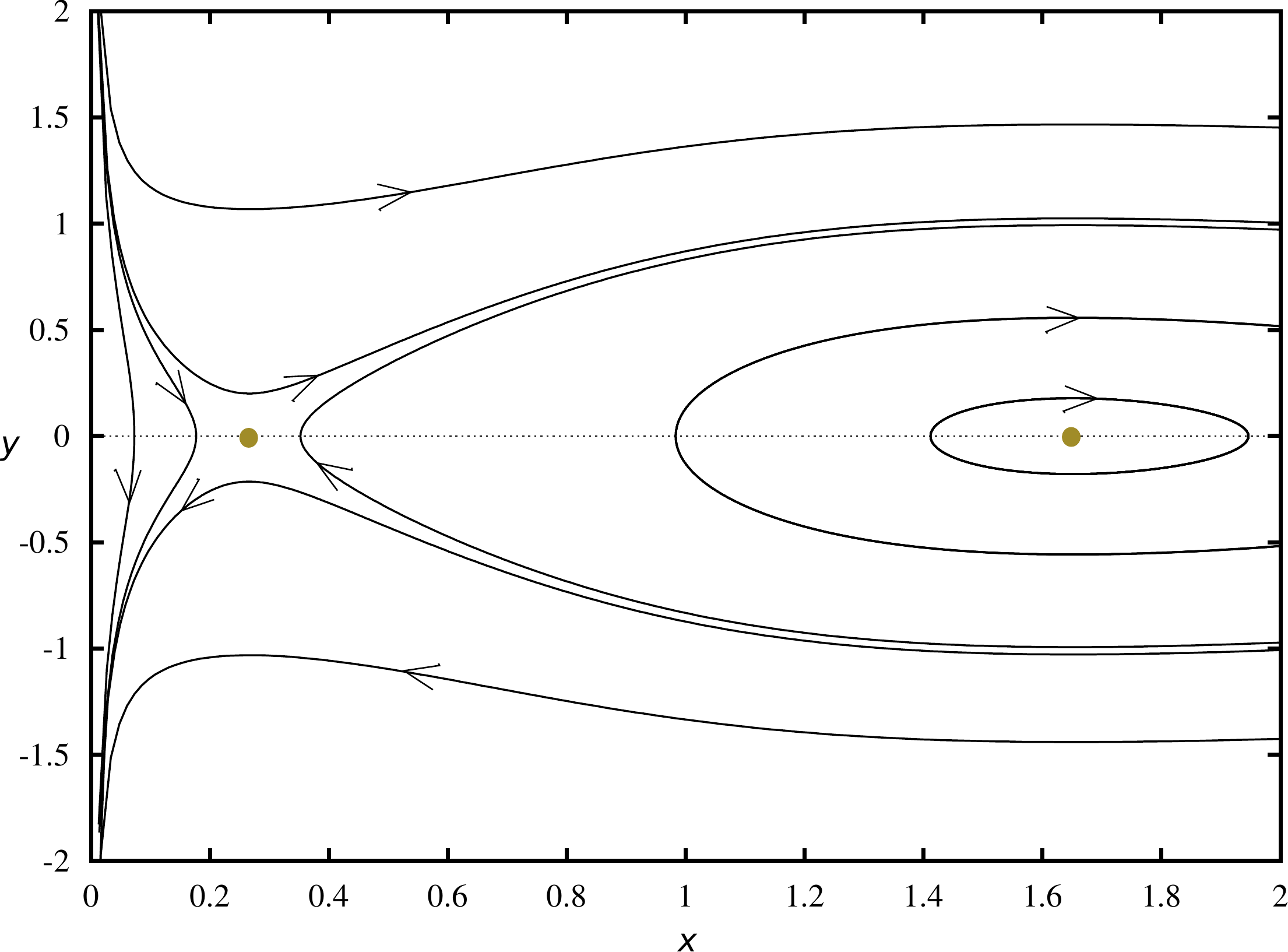}
\caption{The phase portrait of the cosmological model (\ref{eq:27})-(\ref{eq:28}) with $A_s=0.7$, $\Omega_\text{b,0}=0.05$, $\Omega_{x,0}=0.95$ and Chaplygin gas $m=3/2$ ($\alpha = -2$). Trajectories between the saddle and the center are for the accelerating universe. Because of the presence of a non-hyperbolic critical point (the center), this system is structurally unstable. \label{fig:4}}
\end{figure}

In figure~\ref{fig:4} the phase portrait of cosmological model with viscous fluid $m=3/2$ is presented. In both cases the baryonic matter is included in the model ($\Omega_\text{b,0}=0.05$). The absence of baryonic matter ($\Omega_{x,0}=1$) does not change qualitatively the phase portraits in both cases.

In contrast to figure~\ref{fig:3} this system is structurally unstable because of the presence the non-hyperbolic critical point---the center. This means that small perturbation of the right hand of the system (\ref{eq:27})-(\ref{eq:28}) leads to a non-equivalent structure of phase space. For large $x$ the phase space structure is organize by separatrices joining the saddle at finite domain (shown in figure~\ref{fig:4}) and a second saddle on the circle at infinity.

Note that there is also a second argument for the confirmation of the existence of structural instability, namely, there exists the separatrix joining two saddle points on the phase portrait \cite{Perko:2001de}. Therefore, following the Peixoto theorem such systems forms a zero-measure set of systems in the space of dynamical systems on the plane. The model of universe to be structurally stable should possess only the deceleration and acceleration phases \cite{Szydlowski:2006az}.

\section{Duality relation for viscous fluid with $p=\gamma \rho$}

Let us consider now some reflectional symmetries called duality relation \cite{Chimento:2002gb,Aguirregabiria:2003uh,Chimento:2003qy,Dabrowski:2003jm,Aguirregabiria:2004te,Calcagni:2004wu,Szydlowski:2005nu,Chimento:2005au,Chimento:2006gk,Chimento:2005xa,Faraoni:2020tpe} of the model under consideration. The presence of this symmetries give us to generate new solutions from knows due to symmetries.

In the FRW cosmology the basic equation which determines the evolution of the Universe is the acceleration equation
\[
\frac{dH}{dt} = -\frac{1}{2}(\rho_\text{eff}+p_\text{eff})=-\frac{3}{2}(1+w_\text{eff}),
\]
where $\rho_\text{eff}$, $p_\text{eff}$, $w_\text{eff}$ are effective energy density, effective pressure and effective coefficient equation of state.

The first look at this relation unable us to observe reflectional symmetry of the type: $(\rho_\text{eff} + p_\text{eff})$ reflect on $-(\rho_\text{eff}+p_\text{eff})$, and $H$ reflects on $-H$ (the scale factor a change on its inverse $a^{-1}$), then the basic acceleration equation does not change the form, therefore reflections are symmetry operation.

It is a simple interpretation of this symmetry. Let us consider for simplicity case of constant $w_\text{eff}=\gamma$. Then if $a(t)$ is the solution of the acceleration equation with the equation of state parameter $\gamma=-1+ \delta$, then $a^{-1}$ will be for equation of state parameter $(-1 - \delta)$.

It is a simple consequence that $\rho_\text{eff} H^{-2}=\text{const}$ is invariant of the scaling symmetries of the FRW equation \cite{Szydlowski:1983es} (see also \cite{Pailas:2020xhh}).

If we look at a dynamical system with some additional baryonic matter one can observe that this type of symmetry is preserved if $m$ is established. However, let us discover a new type of generalized duality symmetries.

If $(1+\gamma)$ is reflected on a symmetric value with respect to $m=1/2$ on $(1+\gamma)$, then the solution $H(t)$ changes to $H^{-1}$. There is some simple interpretation of this kind duality relation, namely, if $H(t)$ is the solution of the acceleration equation with $\gamma$ equals $(-1+\delta)$ and $m$ equals $(-1/2 +\epsilon)$, then $1/H$ will be also the solution for $\gamma$ equal $-1 - \delta$ and $m$ equal $-1/2 - \epsilon$, where $\delta$, $\epsilon$ are arbitrary constants.

Scaling symmetries called also homological symmetries play an important role in the FRW cosmology \cite{Szydlowski:1983es,Szydlowski:2005nu,Pailas:2020xhh}. In this context Szyd{\l}owski and Heller proved that the FRW dynamical system in state variables $(H,\rho)$ admits a Lie group symmetries \cite[p.~574]{Szydlowski:1983es}
\begin{theorem}
Friedmann's dynamical system
\begin{align*}
\dot{H} &= -H^2 - \frac{1}{6} (\rho + 3p) + \frac{\Lambda}{3} \\
\dot{\rho} &= -3H(\rho + p)
\end{align*}
with $p=\gamma \rho$, $\gamma = \text{const}$, admits a Lie group with the infinitesimal operator $X = - A t \partial/\partial t + A H \partial/\partial H + 2A(\bar{\rho} + \Lambda) \partial/\partial \rho$; and, vice versa, from the invariance of the above system with respect to the operator $X$ it follows the equations of state $p = \gamma \rho + p_0$, $p_0 \sim \Lambda$. The equation of the group invariant is that of the flat model trajectory $\rho -3H^2 + \Lambda = 0$. Finite transformations of the group are: $\bar{H} = He^{\tau}$, $\bar{\rho} + \Lambda = (\rho + \Lambda) e^{2\tau}$, $\bar{t} = te^{-t}$, $(H,\rho) \ne (0,0)$.

\end{theorem}

One can derive basing on these symmetries homological theorems \cite{Collins:1977ss}. From these symmetries $t$ is switch on $t A^{-1}$, $H$ is switch on $AH$ and $\rho$ is switched on $\rho A^2$ is symmetry transformation (for physical interpretation see \cite{Faraoni:2020tpe}). From these theorems we obtain that if $H^2 \rho^{-1} = \text{const}$ is a solution, then $(AH(t))^2(A^2 \rho)^{-1}=\text{const}$ will be also a solution. The homological theorems gives us a simple possibility of generating new solutions from knows ones. If we choose $A=-1$ ($H$ is reflected on $H^{-1}$ or $a$ is reflected on $a^{-1}$), then we obtain a corresponding duality relation. Therefore, homological theorems are good way to find some duality relations.

It is also interesting that the parameter $m=1/2$ is distinguished by homological symmetries. If we assume that the FRW dynamical system (with the term $\Lambda$) with viscous matter admits analogous symmetries generated by the symmetry operator $X$, then symmetries enforce just this value of the viscosity parameter. Therefore, the parameter value ($m=1/2$) distinguished by Brevik \cite{Brevik:2020psp} is also justified by a homological type of symmetries.

\section{Dynamics of two fluids in FRW cosmology}

Let us consider two non-interacting fluids in the FRW cosmology. We assume that one fluid is representing viscous dark matter like in previous single fluid description and second one is perfect fluid in the form of dust matter $p=0$, describing baryonic matter.

The effective potential is addictive, therefore
\[
V_{\text{eff}} = V(\rho_\text{b}) + V(\rho_\text{dm}) = - \frac{a^2}{6} \sum_i \rho_i.
\]

The dynamics of the fluids in FRW cosmology is described, in general, by the three-dimensional dynamical system of the form
\begin{align}
&\frac{dH}{dt} = -H^2 - \frac{1}{6}( \rho_\text{b} + \rho_\text{dm} - 9 \beta \rho^m H), \label{eq:5.1} \\
&\frac{d\rho_\text{b}}{dt} = -3 H \rho_\text{b}, \label{eq:5.2} \\
&\frac{d\rho_\text{dm}}{dt} = -3 H \rho_\text{b} - 3H ( \rho_\text{dm} -3 \beta \rho^m H). \label{eq:5.3}
\end{align}
Dynamical system (\ref{eq:5.1})-(\ref{eq:5.3}) admits the first integral called Friedmann equation in the form
\begin{equation} \label{eq:5.4}
\rho_\text{dm} + \rho_\text{b} = 3H^2+3 \frac{k}{a^2}
\end{equation}
where $k$ is the curvature constant.

For simplicity we assume $k=0$ and consider expanding models, then
\begin{equation} \label{eq:5.5}
H = \left(\frac{\rho_\text{dm} + \rho_\text{b}}{3}\right)^{1/2}.
\end{equation}
Additionally we introduce the Hubble time parameter $\tau$ by parameterizing the original cosmological time parameter $t$
\begin{equation} \label{eq:5.6}
\tau \colon H dt=d \tau = d(\ln a(t)).
\end{equation}
Of course it is a monotonic function of original time $t$ because $H$ is positive.

After substitution relation (\ref{eq:5.5}) into equations (\ref{eq:5.2}) and (\ref{eq:5.3}) and reparameterize the original time $t$ to time $\tau$ (\ref{eq:5.6})
we obtain a two-dimensional dynamical system describing evolution of two fluids, viscous dark matter and pressureless baryonic matter
\begin{align}
\frac{d \rho_\text{dm}}{d \tau} &= -3 \rho_\text{b} - 3\left[ \rho_\text{dm} -3 \beta \rho^m \left(\frac{\rho_\text{dm}+\rho_\text{b}}{3}\right)^{1/2}\right], \label{eq:5.7} \\
\frac{d\rho_\text{b}}{d \tau} &= -3 \rho_\text{b}. \label{eq:5.8}
\end{align}
Let us consider evolution of two fluids in the phase space $(u,v)\colon u=\rho_\text{dm}, v=\rho_\text{b}$. The corresponding system is
\begin{align}
\frac{du}{d \tau} &= -3v -3 \left[ u -3 \beta u^m \left( \frac{u+v}{3} \right)^{1/2} \right], \label{eq:5.9} \\
\frac{dv}{d \tau} &= -3v. \label{eq:5.10}
\end{align}
Singular solutions play important role among the solutions of system (\ref{eq:5.9})-(\ref{eq:5.10}). From the physical point of view they are representing asymptotic states of the system.

There is only one critical point of the system at finite domain of the phase space
\[
u_0 \colon u_0^{m-\frac{1}{2}} = \frac{1}{\sqrt{3}\beta} , \qquad v_0=0.
\]
At this critical point baryonic matter is vanishing and universe is expanding like de-Sitter solution,
\begin{equation} \label{eq:5.11}
H_0=\left(\frac{u_0}{3}\right)^{\frac{1}{2}}.
\end{equation}
Let us study the behavior of trajectories in the phase space in a neighborhood of this critical point.

For this aim let us calculate the linearized system around of this critical point which describes well the original system in the vicinity of this critical point following the Hartmann--Grobman theorem \cite{Perko:2001de}.

The linearization matrix $A$ at the critical point determined the eigenvalues of the linearization matrix satisfying the characteristic equation
\[
\det (A-\lambda \mathbbm{1})=0
\]
where 
\[
\tr A=3 \left(m - \frac{3}{2}\right), \qquad \det A=-9\left( m - \frac{1}{2} \right).
\]

Therefore, for $m<1/2$ we obtain a stable global attractor (stable node) in the form of the de-Sitter solution.
The critical point $\rho_\text{b}=0$ located on $\rho_\text{dm}$ axis is reached by trajectories from its neighborhood regardless of initial conditions. Note that for other barotropic fluids submanifold $\rho_\text{m}=0$ is an invariant submanifold of the general system (\ref{eq:5.1})-(\ref{eq:5.3}) and trajectories are going toward this attracting submanifold. For this system a critical point (non-static) is always an intersection of a boundary of weak energy conditions with a surface representing a flat model. 

\begin{figure}
\includegraphics[width=0.5\textwidth]{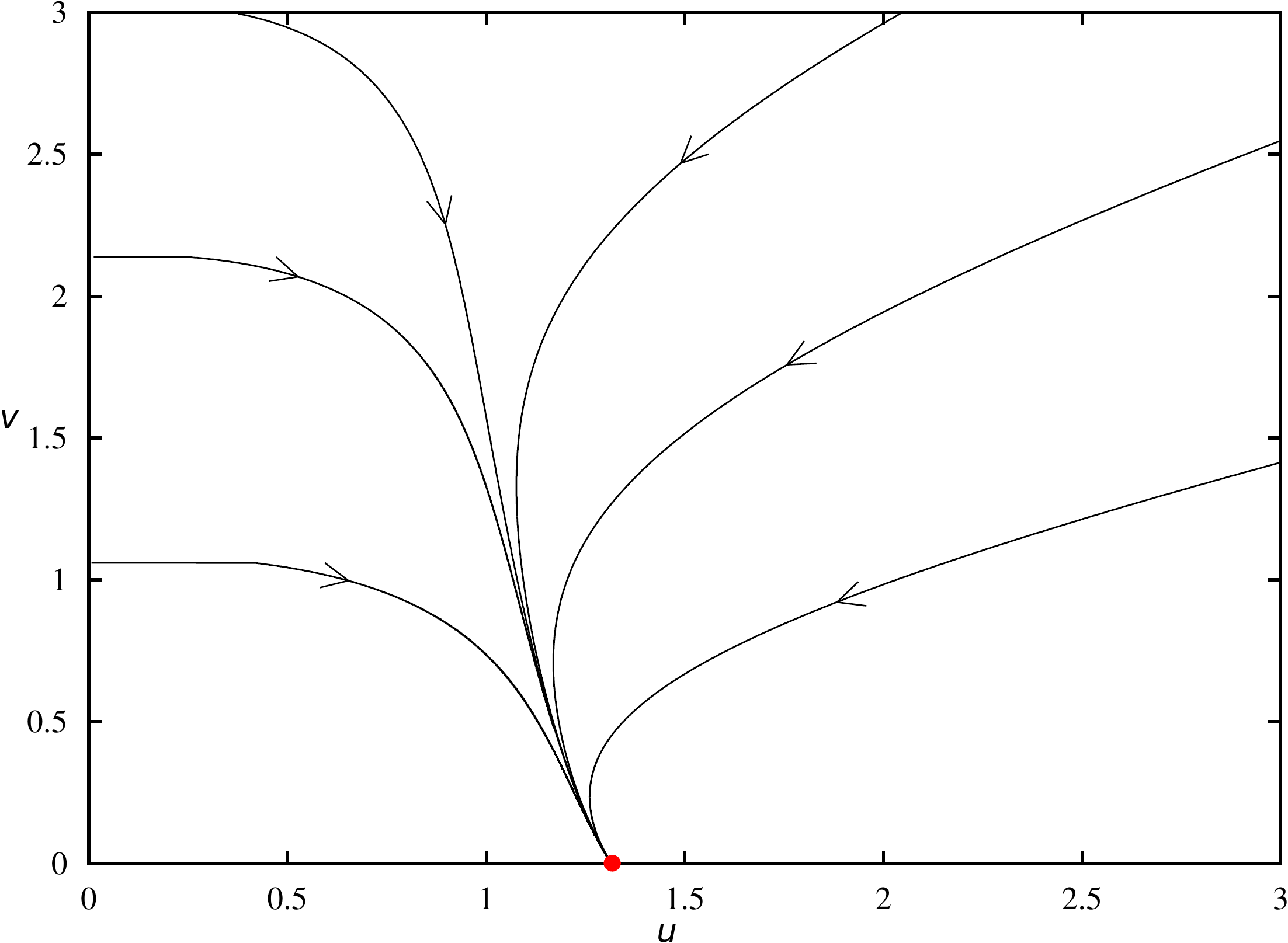}
\caption{The phase portrait of the cosmological model (\ref{eq:5.9})-(\ref{eq:5.10}) in the plane $(u>0 , v \ge 0)$ with viscous matter (Chaplygin gas) and baryonic matter. For illustration it was chosen the value $m = -3/2$ ($\alpha = 1$). The the critical point represents the de Sitter universe (\ref{eq:5.11}) which is a global attractor for all trajectories. Note the universal behavior of the trajectories as they approach the state of single fluid (Chaplygin gas). \label{fig:5}}
\end{figure}

The interval $m < 1/2$ covers the generalized Chaplygin gas case ($-3/2 < m \le -1/2$). The phase portrait for $m=-1/2$ is presented in figure~\ref{fig:5}.

The cosmological model with Chaplygin gas treated as the viscous matter becomes the one fluid model. This validates the approach considered in the previous sections. And, in the long term, the approximation of matter by a single fluid is physically justified (for the late-time viscous cosmology discussion see also \cite{Arora:2020xbn}.

\begin{figure}
\includegraphics[width=0.5\textwidth]{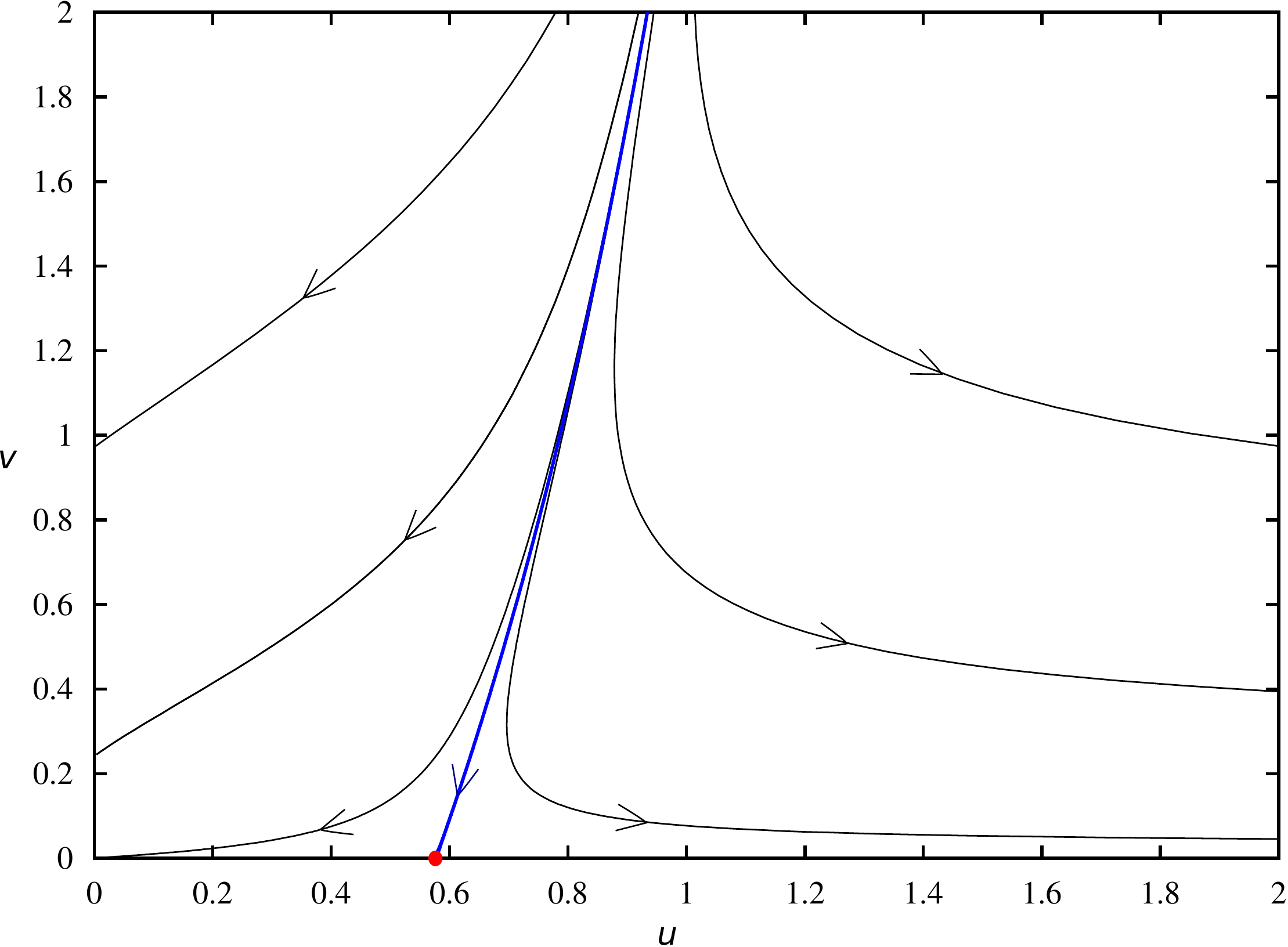}
\caption{The phase portrait of the cosmological model (\ref{eq:5.9})-(\ref{eq:5.10}) in the plane $(u>0 , v \ge 0)$ with viscous matter ($m=3/2$ ($\alpha = -2$)) and baryonic matter. The blue trajectory is the incoming separatrix of the saddle. Note that along this trajectory the system approaches the de Sitter state. \label{fig:6}}
\end{figure}

So far we concentrate on the cosmological model with Chaplygin gas but to make the viscous fluid discussion complete, let us mention the case of $m > 1/2$. The determinant of the linearization matrix $A$ is negative now and the critical point is a saddle. The phase portrait is presented in figure~\ref{fig:6}. There is only one incoming separatrix which reaches the critical point representing the one fluid model.

\section{Conclusion}

Our paper has a theoretical character and concentrate on interpretation of the FRW cosmological model with bulk viscosity effects in the domain of cosmological models with Chaplygin gas. We considered cosmological model with viscous matter without the cosmological constant. Additionally, we make difference between dark matter and visible matter as the former is viscous and the latter non-viscous. In our approach viscosity is rather a attribute of dark matter than a property of environment. The acceleration effect is caused only by viscosity in this class of cosmological models.

We must remember that general relativity is the system with constraints. The energy-momentum tensor gives rise to the FRW equation with dissipation in the Eckart approach \cite{Landau:1987fm}. However, we demonstrated that a corresponding dynamical system with a bulk viscosity effect can be reduced to the dynamical system of a Newtonian type which is a conservative system. Therefore, in investigation of dynamics we can use all methods on conservative nonlinear dynamics (for example classification in the cofiguration space). It is a consequence that bulk viscosity effects does not contribute in the Friedmann first integral (the absence of pressure in the first integral). It is a specific situation which shows explicite that the sense of division on conservative and dissipative dynamical systems in classical mechanics is not applicable for relativistic cosmology.

We reduced the dynamics of dissipative FRW model with bulk viscosity in the Belinskii-Khalatnikov parameterization to the conservative dynamical system of a Newtonian type. We constructed for this system some potential and its dynamics was reduced to the motion of a particle of unit mass moving in this potential which is the function of the scale factor.

Due to this reduction one can classify all possible evolutionary paths directly from the geometry of the potential function. For this aim we considered levels of constant energy (parameterized by the curvature constant parameter). Then the phase portrait can be drawn as in classical mechanics. The potential function is given in the form of a family of two parameter functions of the dimensionless scale factor ($\alpha,\gamma$).

As the $\Lambda$CDM model is considered as a standard one in cosmology, we stress that its phase structure should be inherited by any competitive model. We show that the model with viscous matter ($m < 1/2$) and without the cosmological constant possesses the equivalent structure of phase phase. The equivalence is understood as a homeomorphism of trajectories of both systems preserving the direction of time.

Models with Chaplygin gas are dedicated for realization of unification of dark matter (pressureless, i.e. cold matter satisfying the equation of state for dust, $p=0$) and dark energy given in the form of the cosmological constant parameter, where the form of the equation of state is taken from acoustic considerations \cite{Gorini:2004by}. On the other hand we pointed out on a dissipative origin of this form equation of state in the present paper. In the same sense as quartessence is a way to unify dark matter and dark energy, viscous fluid is an idea of unifying dark energy with dust matter. The framework for this type of unification is the generalized Chaplygin gas.

Moreover, considering the two non-interacting fluids (viscous Chaplygin fluid and non-viscous baryonic matter) we found that Chaplygin fluid dominates in the long term, i.e. one fluid model is an attractor in the phase space of two non-interacting fluid models.

From the astronomical data we know that the $\Lambda$CDM model is the best one in the light of other theoretical hypotheses of dark energy \cite{Szydlowski:2006ay}. After reduction the FRW cosmology dynamics to the FRW dynamics with Chaplygin gas one can draw two-dimensional phase portraits and we get that for reasonable model parameters both portraits for the $\Lambda$CDM model and the CDM model with bulk viscosity are topologically equivalent. Of course, this theoretical requirement puts limits on the values of model parameters.

In the recent paper Brevik and Normann \cite{Brevik:2020psp} have obtained the value of $m= 1/2$ to be distinguished by the current Universe. We have found that cosmological models with viscous matter and without the cosmological term should be structurally stable provided that the parameter $m < 1/2$.

\begin{acknowledgements}
The authors are grateful to the anonymous referee for constructive suggestions and comments on the manuscript.
\end{acknowledgements}

\end{document}